\documentclass[10pt,a4paper,onecolumn]{article}
\usepackage{marginnote}
\usepackage{graphicx}
\usepackage{xcolor}
\usepackage{authblk,etoolbox}
\usepackage{titlesec}
\usepackage{calc}
\usepackage{tikz}
\usepackage{hyperref}
\hypersetup{colorlinks,breaklinks,
            urlcolor=[rgb]{0.0, 0.5, 1.0},
            linkcolor=[rgb]{0.0, 0.5, 1.0}}
\usepackage{caption}
\usepackage{tcolorbox}
\usepackage{amssymb,amsmath}
\usepackage{ifxetex,ifluatex}
\usepackage{seqsplit}
\usepackage{xstring}

\usepackage{float}
\let\origfigure\figure
\let\endorigfigure\endfigure
\renewenvironment{figure}[1][2] {
    \expandafter\origfigure\expandafter[H]
} {
    \endorigfigure
}

\usepackage{fixltx2e} 
\usepackage{natbib}
\let\textttOrig=\texttt
\def\texttt#1{\expandafter\textttOrig{\seqsplit{#1}}}
\renewcommand{\seqinsert}{\ifmmode
  \allowbreak
  \else\penalty6000\hspace{0pt plus 0.02em}\fi}

\makeatletter
\let\href@Orig=\href
\def\href@Urllike#1#2{\href@Orig{#1}{\begingroup
    \def\Url@String{#2}\Url@FormatString
    \endgroup}}
\def\href@Notdoi#1#2{\def\tempa{#1}\def\tempb{#2}%
  \ifx\tempa\tempb\relax\href@Urllike{#1}{#2}\else
  \href@Orig{#1}{#2}\fi}
\def\href#1#2{%
  \IfBeginWith{#1}{https://doi.org}%
  {\href@Urllike{#1}{#2}}{\href@Notdoi{#1}{#2}}}
\makeatother

\usepackage[top=3.5cm, bottom=3cm, right=1.5cm, left=1.0cm,
            headheight=2.2cm, reversemp, includemp, marginparwidth=4.5cm]{geometry}

\titleformat{\section}
  {\normalfont\sffamily\Large\bfseries}
  {}{0pt}{}
\titleformat{\subsection}
  {\normalfont\sffamily\large\bfseries}
  {}{0pt}{}
\titleformat{\subsubsection}
  {\normalfont\sffamily\bfseries}
  {}{0pt}{}
\titleformat*{\paragraph}
  {\sffamily\normalsize}

\usepackage{fancyhdr}
\makeatletter
\let\ps@plain\ps@fancy
\fancyheadoffset[L]{4.5cm}
\fancyfootoffset[L]{4.5cm}

\definecolor{linky}{rgb}{0.0, 0.5, 1.0}

\newtcolorbox{repobox}
   {colback=red, colframe=red!75!black,
     boxrule=0.5pt, arc=2pt, left=6pt, right=6pt, top=3pt, bottom=3pt}

\newcommand{\ExternalLink}{%
   \tikz[x=1.2ex, y=1.2ex, baseline=-0.05ex]{%
       \begin{scope}[x=1ex, y=1ex]
           \clip (-0.1,-0.1)
               --++ (-0, 1.2)
               --++ (0.6, 0)
               --++ (0, -0.6)
               --++ (0.6, 0)
               --++ (0, -1);
           \path[draw,
               line width = 0.5,
               rounded corners=0.5]
               (0,0) rectangle (1,1);
       \end{scope}
       \path[draw, line width = 0.5] (0.5, 0.5)
           -- (1, 1);
       \path[draw, line width = 0.5] (0.6, 1)
           -- (1, 1) -- (1, 0.6);
       }
   }

\patchcmd{\@maketitle}{center}{flushleft}{}{}
\patchcmd{\@maketitle}{center}{flushleft}{}{}
\patchcmd{\@maketitle}{\LARGE}{\LARGE\sffamily}{}{}
\def\maketitle{{%
  
  \AB@maketitle}}
\makeatletter
\renewcommand\AB@affilsepx{ \protect\Affilfont}
\renewcommand\AB@affilnote[1]{{\bfseries #1}\hspace{3pt}}
\renewcommand{\affil}[2][]%
   {\newaffiltrue\let\AB@blk@and\AB@pand
      \if\relax#1\relax\def\AB@note{\AB@thenote}\else\def\AB@note{#1}%
        \setcounter{Maxaffil}{0}\fi
        \begingroup
        \let\href=\href@Orig
        \let\texttt=\textttOrig
        \let\protect\@unexpandable@protect
        \def\thanks{\protect\thanks}\def\footnote{\protect\footnote}%
        \@temptokena=\expandafter{\AB@authors}%
        {\def\\{\protect\\\protect\Affilfont}\xdef\AB@temp{#2}}%
         \xdef\AB@authors{\the\@temptokena\AB@las\AB@au@str
         \protect\\[\affilsep]\protect\Affilfont\AB@temp}%
         \gdef\AB@las{}\gdef\AB@au@str{}%
        {\def\\{, \ignorespaces}\xdef\AB@temp{#2}}%
        \@temptokena=\expandafter{\AB@affillist}%
        \xdef\AB@affillist{\the\@temptokena \AB@affilsep
          \AB@affilnote{\AB@note}\protect\Affilfont\AB@temp}%
      \endgroup
       \let\AB@affilsep\AB@affilsepx
}
\makeatother

\renewcommand\Affilfont{\sffamily\small\mdseries}
\ifnum 0\ifxetex 1\fi\ifluatex 1\fi=0 
  \usepackage[T1]{fontenc}
  \usepackage[utf8]{inputenc}

\else 
  \ifxetex
    \usepackage{mathspec}
  \else
    \usepackage{fontspec}
  \fi
  \defaultfontfeatures{Ligatures=TeX,Scale=MatchLowercase}

\fi
\IfFileExists{upquote.sty}{\usepackage{upquote}}{}
\IfFileExists{microtype.sty}{%
\usepackage{microtype}
\UseMicrotypeSet[protrusion]{basicmath} 
}{}

\usepackage{hyperref}
\hypersetup{unicode=true,
            pdftitle={JAX-bandflux: differentiable supernovae SALT modelling for cosmological analysis on GPUs},
            pdfborder={0 0 0},
            breaklinks=true}
\let\addcontentslineOrig=\addcontentsline
\def\addcontentsline#1#2#3{\bgroup
  \let\texttt=\textttOrig\addcontentslineOrig{#1}{#2}{#3}\egroup}
\let\markbothOrig\markboth
\def\markboth#1#2{\bgroup
  \let\texttt=\textttOrig\markbothOrig{#1}{#2}\egroup}
\let\markrightOrig\markright
\def\markright#1{\bgroup
  \let\texttt=\textttOrig\markrightOrig{#1}\egroup}

\IfFileExists{parskip.sty}{%
\usepackage{parskip}
}{
\setlength{\parindent}{0pt}
\setlength{\parskip}{6pt plus 2pt minus 1pt}
}
\ifx\paragraph\undefined\else
\let\oldparagraph\paragraph
\renewcommand{\paragraph}[1]{\oldparagraph{#1}\mbox{}}
\fi
\ifx\subparagraph\undefined\else
\let\oldsubparagraph\subparagraph
\renewcommand{\subparagraph}[1]{\oldsubparagraph{#1}\mbox{}}
\fi

\title{JAX-bandflux: differentiable supernovae SALT modelling for
cosmological analysis on GPUs}

        \author[1, 2]{Samuel Alan Kossoff Leeney}
    
      \affil[1]{Astrophysics Group, Cavendish Laboratory, J. J. Thomson
Avenue, Cambridge CB3 0HE, UK}
      \affil[2]{Kavli Institute for Cosmology, Madingley Road, Cambridge
CB3 0HA, UK}
  \date{\vspace{-5ex}}

\begin{document}
\maketitle

\marginpar{
  \sffamily\small

  {\bfseries DOI:} \href{https://doi.org/}{\color{linky}{}}

  \vspace{2mm}

  {\bfseries Software}
  \begin{itemize}
    \setlength\itemsep{0em}
    \item \href{}{\color{linky}{Review}} \ExternalLink
    \item \href{}{\color{linky}{Repository}} \ExternalLink
    \item \href{}{\color{linky}{Archive}} \ExternalLink
  \end{itemize}

  \vspace{2mm}

  {\bfseries Submitted:} \\
  {\bfseries Published:} 

  \vspace{2mm}
  {\bfseries License}\\
  Authors of papers retain copyright and release the work under a Creative Commons Attribution 4.0 International License (\href{https://creativecommons.org/licenses/by/4.0/}{\color{linky}{CC BY 4.0}}).
}

\section{Summary}\label{summary}

\href{https://github.com/samleeney/JAX-bandflux}{JAX-bandflux} is a JAX
\citep{jax2018github} implementation of critical supernova modelling
functionality for cosmological analysis. The codebase implements key
components of the established library SNCosmo \citep{barbary2016sncosmo}
in a differentiable framework, offering efficient parallelisation and
gradient-based optimisation capabilities through GPU acceleration. The
package facilitates differentiable computation of supernova light curve
measurements, supporting the inference of SALT \citep{kenworthy2021salt3,
pierel2022salt3} parameters necessary for cosmological analysis.

\section{Statement of need}\label{statement-of-need}

Accurate estimation of supernova flux is essential in cosmological
studies. These measurements are fundamental to the calibration of
standard candles and subsequent distance determinations, which are used
to answer cosmological questions. For example, the rate of expansion of
the universe. Current packages such as SNCosmo \citep{barbary2016sncosmo}
are widely used for analysing supernova data. However, traditional
implementations are not designed to run on GPUs and they lack
differentiability. A differentiable approach enables efficient gradient
propagation during parameter optimisation and supports large-scale
parallel computations on modern hardware such as GPUs. This JAX
implementation addresses these requirements by providing differentiable,
parallelisable routines for SALT parameter extraction.

\section{Implementation}\label{implementation}

The package is structured into several modules and example scripts that
demonstrate various aspects of the supernova modelling workflow. Two
primary example scripts, \texttt{fmin\_bfgs.py} and \texttt{ns.py},
illustrate optimisation via L-BFGS-B and nested sampling respectively.
These scripts utilise core routines from the JAX modules, following a
structure similar to SNCosmo while enabling differentiability and GPU
acceleration. The central computation is contained in the file
\texttt{salt3.py}, which implements the SALT3 model.

The SALT model is of the form: \[
F(p, \lambda) = x_0 \left[ M_0(p, \lambda) + x_1 M_1(p, \lambda) + \ldots \right] \times \exp \left[ c \times CL(\lambda) \right]
\] where free parameters are: \(x_0\), \(x_1\), \(t_0\), and \(c\).
Model surface parameters are: \(M_0(p, \lambda)\) and
\(M_1(p, \lambda)\) are functions that describe the underlying flux
surfaces, and \(p\) is a function of redshift and \(t-2\).

The computation of the bandflux is achieved by integrating the model
flux across the applied bandpass filters. Combining multiple bands, the
bandflux is defined as: \[
\text{bandflux} = \int_{\lambda_\text{min}}^{\lambda_\text{max}} F(\lambda) \cdot T(\lambda) \cdot \frac{\lambda}{hc} \, d\lambda
\] Here, \(T(\lambda)\) is the transmission function specific to the
bandpass filter used; \(h\) and \(c\) are the Planck constant and the
speed of light respectively.

Within \texttt{salt3.py}, the implementation computes the rest-frame
model flux by combining the base spectral surface \(M_0(p, \lambda)\)
with the stretch-modulated variation \(M_1(p, \lambda)\), each scaled by
their respective SALT parameters. These operations utilise JAX's
vectorised array manipulations, which are JIT-compiled for efficient,
parallel execution on GPUs. The resulting flux is computed in a fully
differentiable manner. The computed flux is then multiplied by the
instrument's transmission function \(T(\lambda)\) and by the wavelength
factor \(\lambda/(hc)\), followed by trapezoidal integration along the
wavelength dimension using JAX's numerical integration capabilities.
These operations are also JIT-compiled and can be parallelised across
multiple data instances via \texttt{vmap}.

The package includes comprehensive bandpass filter handling through the
\texttt{bandpasses.py} module, which provides a \texttt{Bandpass} class
to represent filter transmission functions. A set of commonly used
astronomical filters is pre-integrated into the system, whilst
additional custom bandpasses can be registered as needed through
functions such as \texttt{register\_bandpass} and
\texttt{load\_bandpass\_from\_file}. The system also facilitates the
creation of bandpass objects from the Spanish Virtual Observatory (SVO)
filter service. For data handling, the \texttt{data.py} module offers
utilities for loading and processing supernova observations from various
formats, including functions to handle redshift data and prepare it for
model fitting. The package currently supports both SALT3 and SALT3-NIR
models through dedicated interpolation routines found in
\texttt{salt3.py}.

This architecture allows gradient propagation through the entire
analysis pipeline, enabling techniques that benefit from JAX's
differentiable, parallelisable programming paradigm. The implementation
maintains functional parity with SNCosmo whilst providing an enhanced
computational efficiency and scalability for contemporary cosmological
analyses.

\bibliography{paper.bib}

\end{document}